\documentclass[12pt,A4,twoside]{article}
\usepackage[latin1]{inputenc}
\usepackage{amssymb}
\usepackage{amsmath}
\usepackage{amsfonts}
\usepackage{amsbsy}
\usepackage{natbib}
\usepackage{eucal}
\usepackage{graphicx}
\usepackage{titlesec}
\usepackage[english]{babel}
\usepackage{times}
\usepackage{titling}
\usepackage{mathptmx}
\usepackage{float}
\usepackage[font=small]{caption}
\usepackage{array}
\usepackage[colorlinks,citecolor=blue,urlcolor=blue]{hyperref}
\usepackage{authblk}
\usepackage{amsthm}
\usepackage{bm}

\usepackage{setspace}
\usepackage{soul}

\usepackage{cancel}

\DeclareSymbolFont{txgreek}{OML}{cmr}{m}{it}

\renewcommand{\abstract}[1]{{\small\noindent
\hrulefill\par \vspace*{0.1cm}\noindent{\small\bf\sffamily
{Abstract}}\parindent=0pt\par\noindent\vspace{-0.1cm}\noindent\hrulefill\par\vspace*{0.5\baselineskip}\hspace*{0cm}\renewcommand{\baselinestretch}{1.1}\sffamily{#1}\par\vspace*{-0.1cm}\noindent\hrulefill}}

\def\and{,\;}
\DeclareMathSizes{11}{11}{7.8}{6.5}

\def\paragraf{\fontsize{9}{10pt}\fontfamily{phv}\fontshape{it}\selectfont}
\titleformat{\paragraph}
{\titlerule[0pt]
\vspace{0cm}%
\paragraf}
{\theparagraph}{.5em}{\vspace*{0\baselineskip}}

\def\titol{\fontsize{12.045}{12pt}\fontfamily{phv}\fontseries{b}\selectfont}
\titleformat{\section}
{\titlerule[0pt]
\vspace{0.2cm}%
\titol}
{\thesection.}{.5em}{\vspace*{0\baselineskip}}

\def\titolp{\fontsize{11.045}{11pt}\fontfamily{phv}\fontseries{b}\fontshape{it}\selectfont}
\titleformat{\subsection}
{\titlerule[0pt]\bigskip
\vspace{-0.4cm}%
\titolp}
{\thesubsection.}{.5em}{\vspace*{0\baselineskip}}

\def\titolpp{\fontsize{10.045}{10pt}\fontfamily{phv}\fontshape{it}\selectfont}
\titleformat{\subsubsection}
{\titlerule[0pt]
\vspace{-0.4cm}%
\titolpp}
{\thesubsubsection.}{.5em}{\vspace*{0\baselineskip}}

\usepackage{titling}
    \pretitle{\begin{center}\sffamily\fontsize{18pt}{20pt}\selectfont}
    \posttitle{\par\end{center}}
    \preauthor{\begin{center}\fontsize{12pt}{14pt}\selectfont}
    \postauthor{\par\end{center}\vspace{0bp}}
    \predate{}
    \date{}
    \postdate{}

\title{Bayesian hierarchical nonlinear modelling of intra-abdominal volume during pneumoperitoneum for laparoscopic surgery}

\thanksmarkseries{arabic}
\author{Gabriel Calvo\thanks{Department of Statistics and Operations Research, Universitat de Val\`encia, Carrer Doctor Moliner 50, 46100, Burjassot, Spain.}\  \and  Carmen Armero$^1$\and Virgilio G\'omez-Rubio\thanks{Departamento de Matem\'aticas, Escuela T\'ecnica Superior de Ingenieros Industriales, Universidad de Castilla-La Mancha, Avda, de Espa\~na s/n, 02071, Albacete, Spain.} \and Guido Mazzinari\thanks{Research Group in Perioperative Medicine and Department of
Anaesthesiology, Hospital Universitari  i Polit\`ecnic  la Fe,   Avinguda de Fernando Abril Martorell  106, 46026, Val\`encia, Spain}}
\def\headers#1{\fontsize{8.5}{10}\centering\sffamily\itshape{#1}}
\def\page#1{\fontsize{8.5}{10}\sffamily{#1}}
\usepackage{fancyhdr}%Headers

\begin{document}
\maketitle

% Headers
\thispagestyle{empty}
\renewcommand{\headrulewidth}{0truecm}
\pagestyle{fancy}
\rhead[\headers{Bayesian hierarchical nonlinear  modelling  for laparoscopic surgery}]{\page{\thepage}}
\lhead[\page{\thepage}]{\headers{G. Calvo et al.}}
 \lfoot{} \rfoot{}
\cfoot{}

\abstract{Laparoscopy is an operation carried out in the abdomen or pelvis through small incisions with external visual control by a camera. This technique needs the abdomen to be insufflated with carbon dioxide to obtain a working space for surgical instruments' manipulation. Identifying the critical point at which insufflation should be limited is crucial to maximizing surgical working space and minimizing injurious  effects. Bayesian nonlinear growth mixed-effects models are applied to data coming from a repeated measures design. This study allows  to assess the relationship between the insufflation pressure  and the intra--abdominal volume. }

\paragraph{MSC: 62P10, 62F25. }

\paragraph{Keywords: Intra-abdominal pressure,  logistic growth function, Markov chain Monte Carlo methods, random effects.}

\renewcommand{\baselinestretch}{1.2}
\bigskip

\section{Introduction}\label{sec1}

Laparoscopy is an operation carried out in the abdomen or pelvis through small incisions with the help of a camera. It is performed by insufflating $CO_2$  into the abdomen that yields a working space, i.e., pneumoperitoneum,  and passing surgical  instruments through small incisions using a camera to have external visual control of the procedure \citep{Neugebauer}.  Laparoscopy has been gaining ground since its inception because it is associated with less morbidity than the traditional method performed through a single, larger skin incision \citep{Pache}.

The introduction of $CO_2$ into the abdomen is performed by medical devices, i.e., laparoscopic insufflators, through small plastic tubes, i.e. trocars, inserted in the patient's abdominal wall. Laparoscopy technological development has been limited to improvements in camera image quality, whereas little innovation has been made in insufflation devices \citep{Colon}.

The $CO_2$ insufflation pressure, i.e., intra--abdominal pressure ($IAP$), is set manually on the insufflator by the surgical team.  $IAP$ is measured in millimeters of mercury (mmHg), and the usual figures during laparoscopic surgery range between 12 and 15 mmHg. Although international guidelines recommend working with the lowest $IAP$ value that ensures an adequate working space, the standard practice is still to initially set the $IAP$ value without further adjustments regardless of the amount of generated intra--abdominal volume ($IAV$) \citep{Neudecker}, measured in litres (L). Operating at such high $IAP$ increases perioperative morbidity since it leads to decrease abdominal blood perfusion, greater postoperative pain, peritoneal injury, and increased risk of pulmonary complications.

The abdominal compartment shows an anisotropic behavior during pneumoperitoneum which is explained by its combination of rigid borders, e.g., spine, rib cage, and pelvis, and semirigid borders, e.g., abdominal wall muscles and the diaphragm \citep{Becker}. Initially, marginal gains in volume in response to pressure increments are proportional. In other words, the abdominal compliance (C$_{abd}$) which defines the change in volume determined by a change in pressure, follows an approximately linear relationship \citep{Mulier}. According to biomechanics laws, the yield stress point is eventually reached, after which applying additional pressure leads to diminishing gains in volume  \citep{Forstemann}. Identifying this critical point at which insufflation should be limited is crucial to maximizing surgical working space while minimizing injurious $IAP$ effects.

The abdomen pressure--volume dynamics during pneumoperitoneum has been discussed in previous papers \citep{Diaz1, Diaz2, Mazzinari1, Mazzinari2}. These studies suggest   the adequacy of  an increasing sigmoidal model for describing the relationship between both variables.  The aim of this paper is to estimate such a model to gain knowledge about the relationship between $IAP$  and $IAV$, especially about the parameters that determine the different growth stages of the process  in accordance with the specific characteristics of the individuals in the target population.  The hypothesis is that, in a personalised medicine environment,  patient responses to insufflation can be estimated and predicted so that an ideal $IAP$ value could be determined to optimise $IAV$ with the lowest risks of potential negative effects.

The statistical framework of this study are nonlinear growth mixed-effects models, also known as hierarchical nonlinear growth models. They have a long and important scientific tradition  for describing biological, medical, and environmental growth phenomena such as pharmacokinetics \citep{Giltinan}, epidemiology \citep{Lindsey}, physiological-response processes \citep{Peek}, or  forestry \citep{Fang} among others. One of the major appeals of these models is that their parameters contain direct and intuitive information on  the process under study. This fact generates a multifaceted knowledge about the phenomena in question of great scientific value \citep{Davidian}.

Data for the study come from a repeated measures design \citep{Lindstrom}. In our case, the   variable of interest $IAV$ is measured for each individual with regard to different $IAP$ values. This design generates two types of data:  data from the same individual and data from several individuals. Random effects in these models are essential elements to glue   together the different observations of the same individual as they could be considered as a within-individual  variation \citep{Laird}.

The statistical analysis of the problem has been carried out using Bayesian inference. This statistical  methodology accounts for uncertainty in terms of probability distributions
\citep{Loredo1, Loredo2} and uses Bayes' theorem to update all relevant information.
Bayesian statistics  allows to draw individual's inferences   and population outcomes. This feature of Bayesian models is of utmost importance in the case of growth models because it   expresses  in a  natural probabilistic way   all information about the parameters and other relevant features of the growth process through the respective posterior distribution.

 The paper is organised as follows. Section \ref{sec2} presents the data and contains a brief description of them   that  emphasises   the particular features of the repeated-means design through
 the number of observations per individual and their $IAV$ trajectories  according to $AIP$ values. Section \ref{sec3} introduces and formulates the statistical modelling. Subsection \ref{subsec31} discusses  the posterior distribution of the inferential process. Subsections \ref{subsec32} and \ref{subsec33} contain, respectively, some relevant results of clinical interest at specific individual levels and in general terms for different population groups. The paper ends with an overview of the results and some conclusions.

\section{Intra-abdominal volume and intra-abdominal pressure data}\label{sec2}

The data for the current modelling come from a previously published individual patient meta--analysis \citep{Mazzinari2} that included experimental information from three previous clinical studies \citep{Mazzinari1, Diaz1, Diaz2}. All patients in these studies underwent a standardized pneumoperitoneum insufflation at a constant low flow, i.e., 3 L$\,$min$^{-1}$, under deep neuromuscular block with a posttetanic count ($PTC$) between one and five assessed by quantitative monitoring. The insufflation was carried out through a leakproof trocar up to an $IAP$ of 15mmHg  for abdominal wall prestretching and then stepwise changes in $IAP$ in the 8 to 15 mmHg pressure range were recorded.  In all studies, patients' legs were placed in padded leg-holder supports with hips flexed before the initial insufflation.

\begin{figure}[h]
\begin{center}
\includegraphics[width= 9cm]{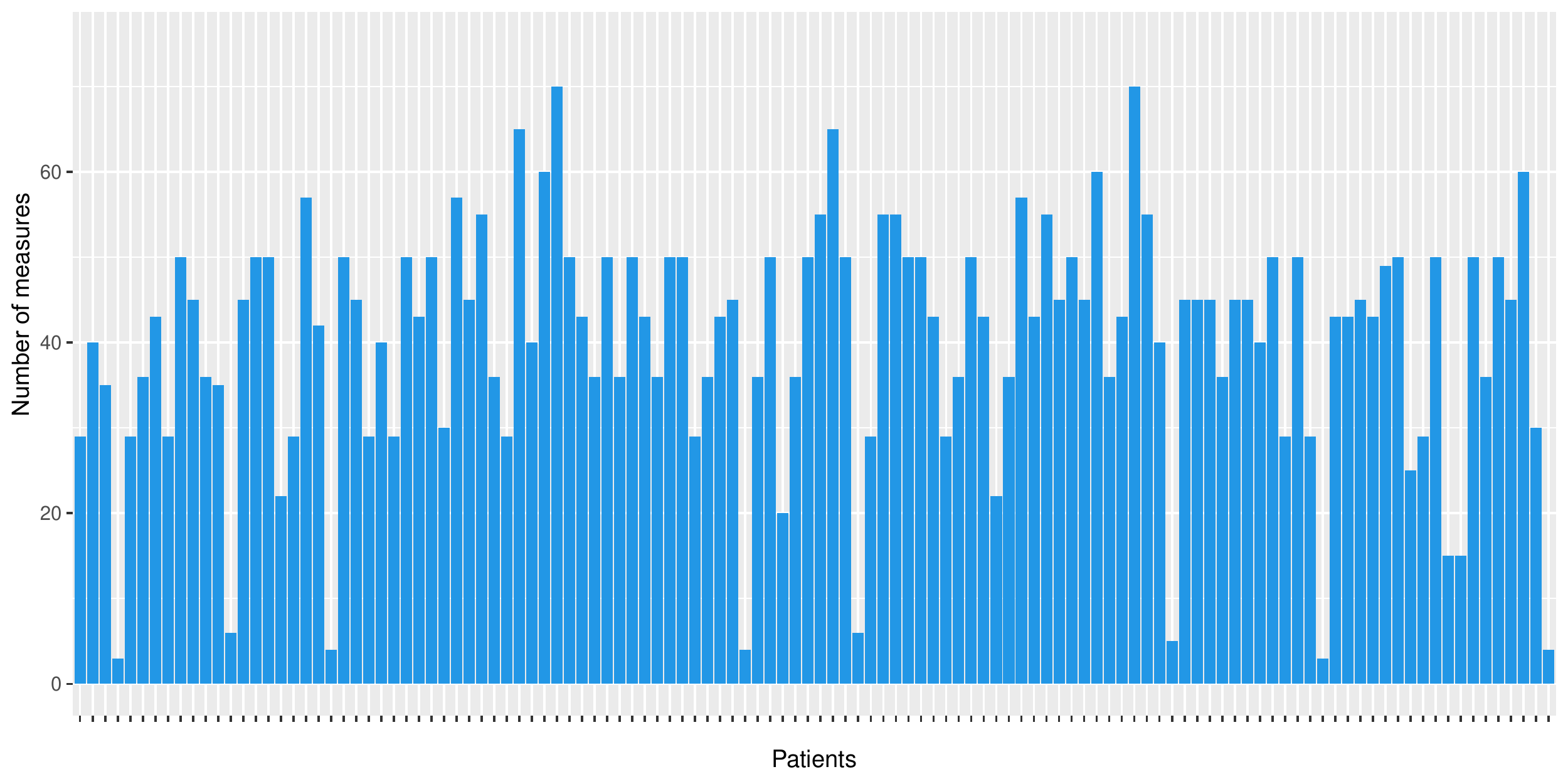}\\
\includegraphics[width= 9cm]{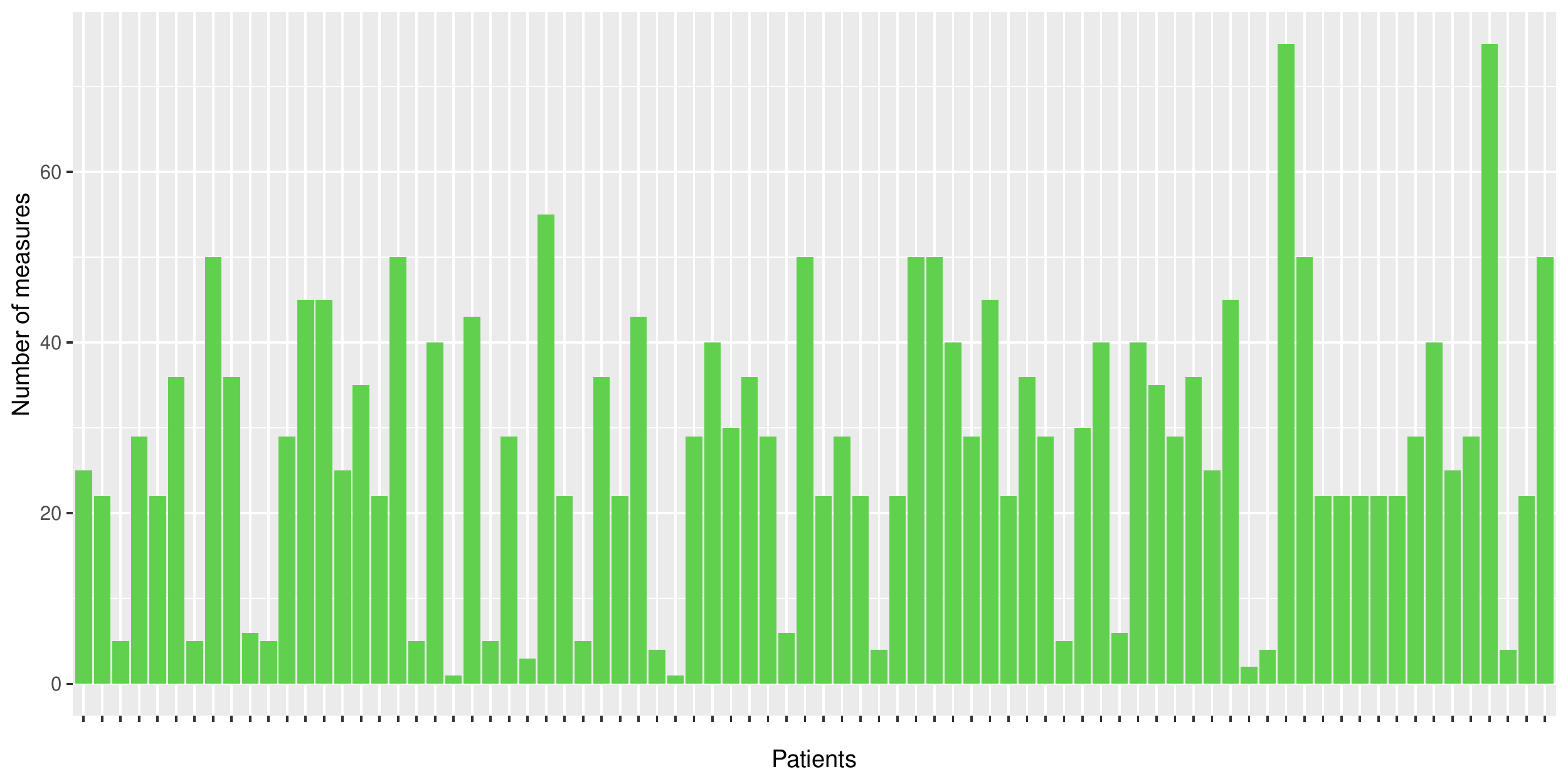}\\
\caption{Number of repeated measures in the men's group (top panel) and in the women's group (bottom panel). Each bar corresponds to a person and its ordinate describes the number of visits of that person during the study. Patients are ordered according to their age from youngest to oldest.  } \label{fig:repeteaded}
\end{center}
\end{figure}

The original databank had information on 204 patients, but 6 patients presented missing information on $IAP$, $IAV$, and/or age values. There are very few individuals whose missing observations do not appear to have been generated    by non-ignorable mechanisms. For this reason, we decided to eliminate them directly and not engage in a very unhelpful imputation process.  The final databank has 198 patients, 118 men and 80 women, and a total of
6$\hspace*{0.04cm}$985 observations. We have a repeated measures design with a very different number of  observations per individual: from individuals with only one observation to individuals with 75. Figure~\ref{fig:repeteaded} shows the number of repeated measures for the group of men and of women  in order of age. It is interesting to note that women
 have less measurements than men in all ages, but especially when they are young.

The data have a very wide age range. The youngest patient is 23 years old and the oldest is 92, with a mean age of 64.65 years. In the men's group, the minimum and maximum also are 23 and 92, respectively, and their average is 64.49 years. Women have a minimum age of 34.77 and a maximum of 85.92, and their mean is 64.87 years.
%Figure~\ref{fig:violin} is a violin plot of the ages of the men and of the women that clearly visualizes the age distribution in both groups.

%\begin{figure}[h]
%\begin{center}
%\includegraphics[width= 7cm]{img/violin_plot_age_sex.pdf}
%\caption{Violin plot of the age of the women and men in the sample. } \label{fig:violin}
%\end{center}
%\end{figure}

\begin{figure}[h]
\begin{center}
\includegraphics[width= 9cm]{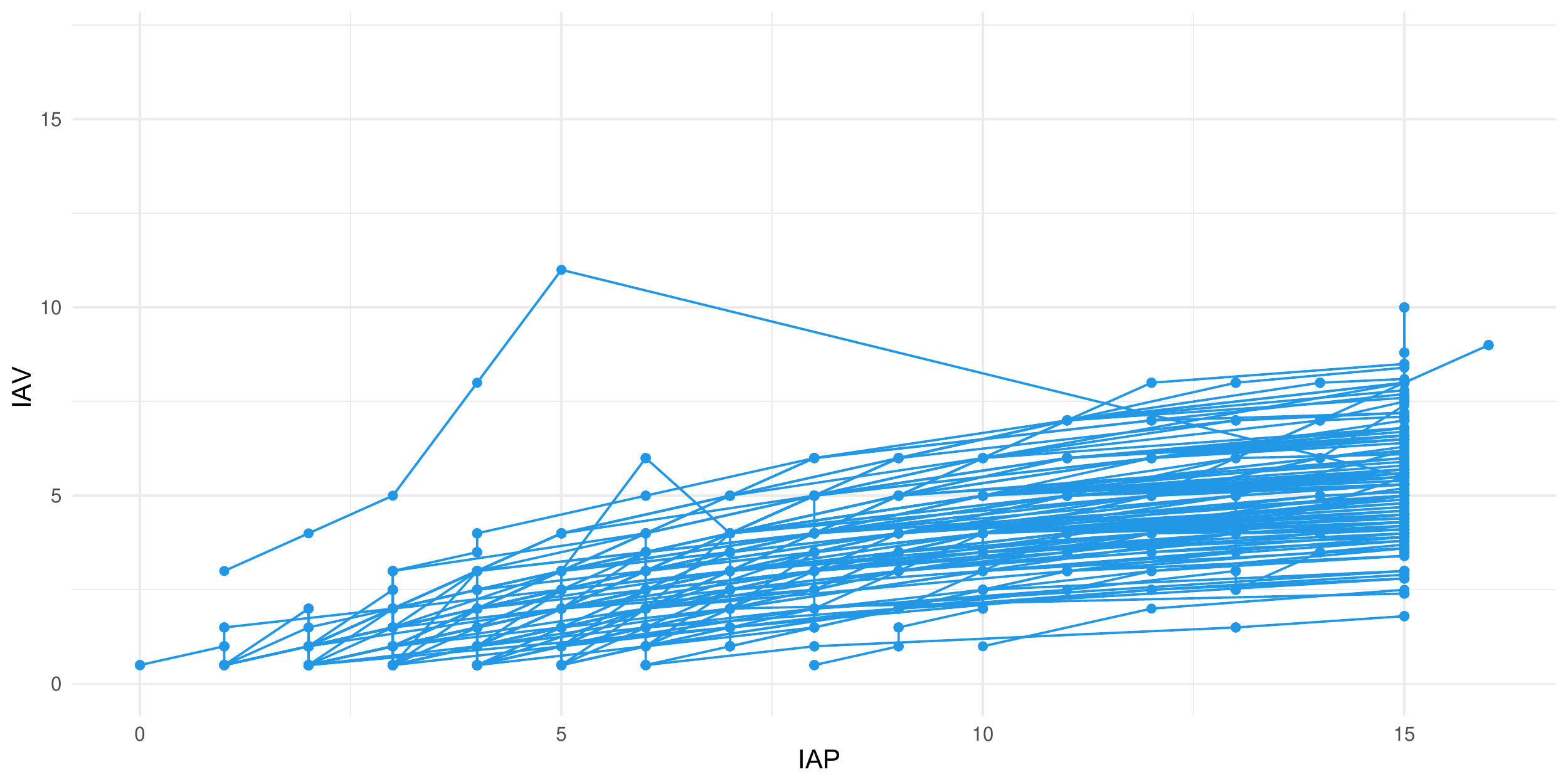}\\
\includegraphics[width= 9cm]{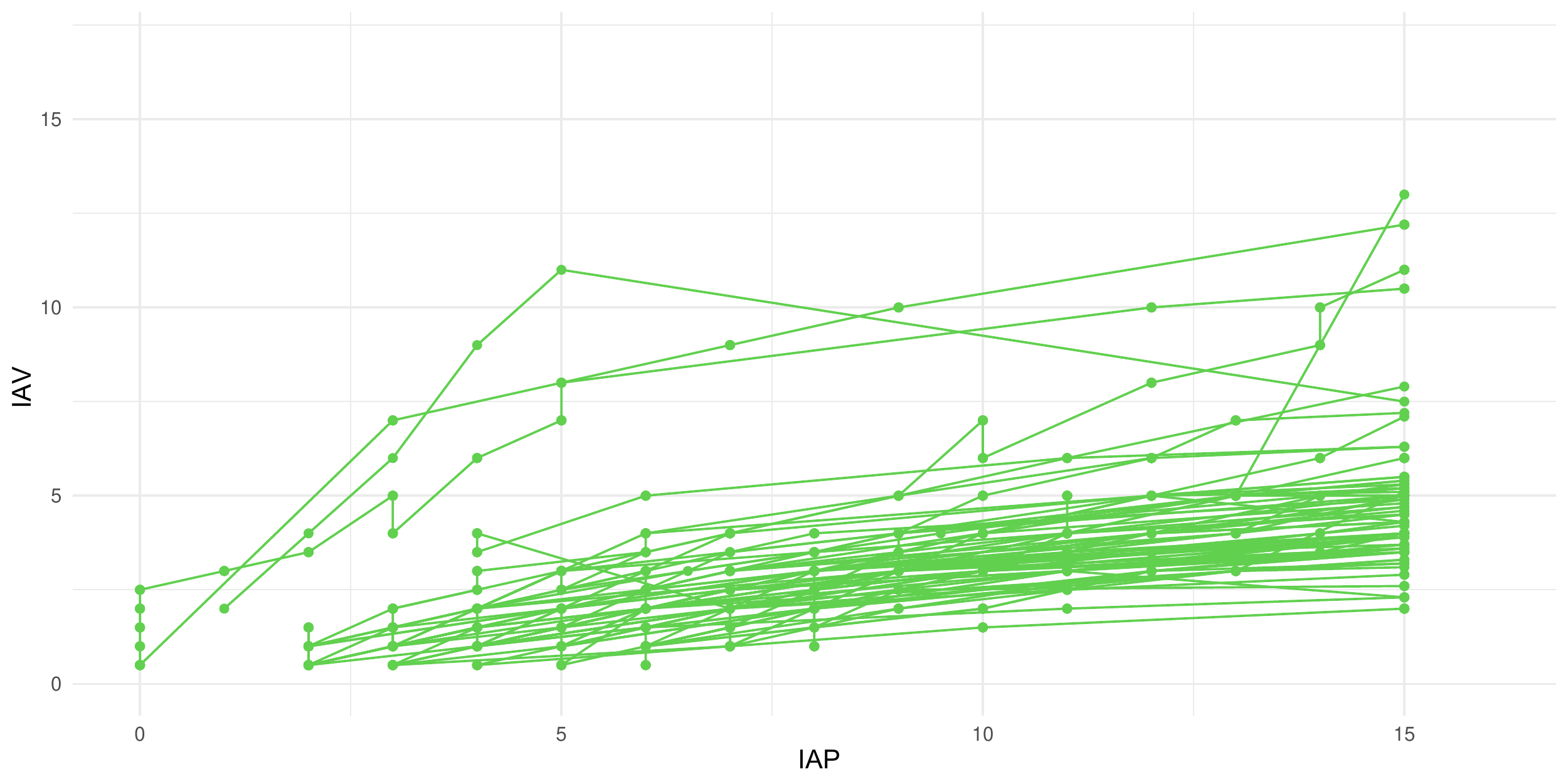}
\caption{$IAV$  profiles (in L) according to $IAP$  (in mmHg) for men (top panel) and women (bottom  panel) in the sample.}\label{fig:spaghetti}
\end{center}
\end{figure}

$IAP$  values range between 0 and 16 mmHg, and $IAV$ values between 0.5 and 13 L.  Figure~\ref{fig:spaghetti} shows a spaghetti plot of $IAV$, in L, for men and women. Men and women show a fairly similar pattern of the $IAV$ with $IAP$, although a greater range of values is observed in men, especially in large values of $IAP$.  In both groups there are individuals with different behaviour but men behave more homogeneously than women.

\section{Logistic growth  mixed-effects modelling}\label{sec3}

Let the  nonlinear mixed-effects model  for the random variable $Y_{ij}$  that records the $IAV$  value for individual $i$, $i=1,\ldots,n$ with   standardized  $IAP$  value $x_{ij}$, $j=1,\ldots, J_i$, defined  in terms of a conditional normal distribution as follows
\begin{equation}\label{eqn:modela}
(Y_{ij} \mid  \mu_{ij}, \sigma^2) \sim \mbox{N}(\mu_{ij}, \sigma^2),
\end{equation}

\noindent where the mean $\mu_{ij}$ is the true $IAV$ value of a patient with  $IAP$ value $x_{ij}$ and can  be expressed in terms of the conditional logistic growth function
\begin{equation}\label{eqn:modelb}
 (\mu_{ij} \mid   a_i, b_i, c_i, x_{ij}) = \frac{a_i}{1+\mbox{exp}\{-(b_i+c_i \, x_{ij})\}},
\end{equation}

\noindent with parameters  $a_i$, $b_i$, and $c_i$    determining  the growth of the function, and $\sigma^2$  the unknown variance associated to the random measurement error of the normal (\ref{eqn:modela}).

\begin{figure}[H]
\begin{center}
\includegraphics[width= 11cm, height=5cm]{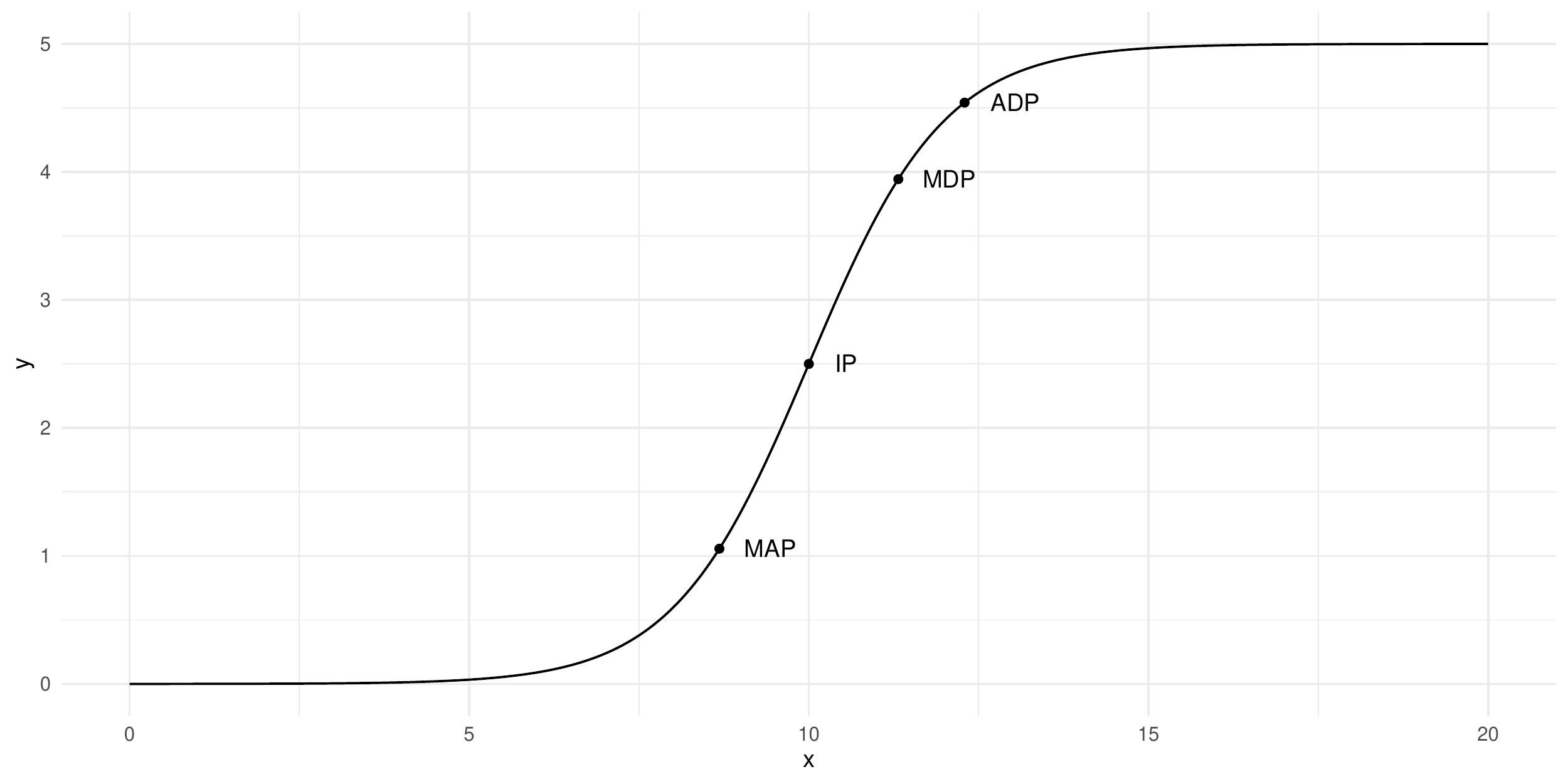}
\caption{Graphics of the logistic growth function   $5/[1+ \mbox{exp}\{-(-10+x) \}]^{-1}$, the subsequent  asymptotic value, and its $MAP$, $IP$, $ADP$, and $MDP$ points.} \label{fig:logistic}
\end{center}
\end{figure}

The logistic growth model for $\mu_{ij}$ has important features which are very valuable  to better understand the relationship between $IAP$ and $IAV$ \citep{Davidian}:

 \begin{itemize}
 \item It is an increasing sigmoid function (see Figure~\ref{fig:logistic}), or $S$-curve, whose name comes from its shape and   was   introduced by the mathematician Pierre-Fran\c{c}ois Verhulst  in the 19th century to study the growth of populations in autocatalytic chemical reactions \citep{Cramer}.
 \item The  asymptotic value of $\mu_{ij}$ when $IAP$ goes to infinity is $a_i$.
 \item The inflection point ($IP$),  where the curve  changes from being  concave downward  to  concave upward and therefore it is the point at which the acceleration of the process switches from positive to negative, is
$-b_i/c_i$. The   value of $\mu_{ij}$ at this point is $a_i/2$.
\item The asymptotic deceleration point ($ADP$), which determines the point   from which the deceleration of the function is very slow  and  it is expected, therefore, that the increase of the function is not of much practical practical interest, is $-(\mbox{ln}(5-2 \sqrt{6})+b_i)/c_i$. The value of $\mu_{ij}$ at this point is  $a_i(3+\sqrt{6})/6$.
    \item The maximum acceleration and  deceleration point, $MAP$ and $MDP$ respectively, and the subsequent true $IAV$ value is      $((-(\mbox{ln}(2+\sqrt{3})+b_i)/c_i,\,   a_i/(3-\sqrt{3}))$ and $(-(-\mbox{ln}(2-\sqrt{3})+b_i)/c_i,\,  a_i/(3+\sqrt{3}))$.
\end{itemize}

By way of illustration, Figure (\ref{fig:logistic}) shows the graph of the logistic growth model $y=5/[1+ \mbox{exp}\{-(-10+x) \}]^{-1}$ and the location on the graph of the special points described above.

Hierarchical modelling for parameters $a_i$, $b_i$ was based on expert information and connected them with covariates age and gender. Parameter $c_i$ was associated to covariate gender. We discarded its connection to covariate age as a consequence of a previous analysis of variable selection that we will discuss later. Furthermore, $a_i$ and $b_i$   also included a random effect specifically associated to each individual that allow to connect all their repeated observations. We have not included any random effect in the modelling of the parameter $c_i$ because it would generate a random interaction term with the $IAP$ values that would be difficult to understand and justify. Following this reasoning, our model would be

%Expert information connects   parameters $a_i$ and $b_i$ to covariates age and gender. There does not seem to be a definitive view on the relationship of $c_i$ to both covariates. In this context, we   considered different modelling approaches  for $c_i$, without the two covariates, with only one, and with both.  The Deviance Information Criterion \citep{Spiegelhalter} was used for  model comparison. According to this criterion  the best  model was the one with only the gender covariate and a common population term in parameter $c_i$.
	 \begin{align}
	     a_i &= \beta_0^{(a)} + u_{i}^{(a)} + \beta_W^{(a)} I_{W}(i)  + \beta_A^{(a)} Age_i,\\
	     b_i &= \beta_0^{(b)} + u_{i}^{(b)} + \beta_W^{(b)} I_{W}(i)  + \beta_A^{(b)} Age_i,\\
         c_i &= \beta_0^{(c)} +   \beta_W^{(c)} I_{W}(i),
	 \end{align}
\noindent where $\bm \beta_0 =(\beta_0^{(a)}, \beta_0^{(b)}, \beta_0^{(c)})^\prime$ stands for the common intercept with the men group being the reference group, $I_{W}(i)$ is the indicator variable   with value 1 if individual $i$ is a woman and
0 otherwise, $\bm \beta_W=(\beta_W^{(a)}, \beta_W^{(b)}, \beta_W^{(c)})^\prime$ and
 $\boldsymbol \beta_A =(\beta_A^{(a)}, \beta_A^{(b)})^\prime$ are the vector of regression coefficients associated with individual $i$   being a woman and their  standardized  age,  respectively.
  Random effects $u_{i}^{(a)}$ and $u_{i}^{(b)}$, $i=1,\ldots,n$, are   assumed conditional independent given $\sigma^2_a$ and  $\sigma^2_b$  and  normally distributed according to  $f(u_{i}^{(a)}|\sigma_{a}^2)= \text{N}(0, \sigma_{a}^2)$ and $f(u_{i}^{(b)}|\sigma_{b}^2)= \text{N}(0, \sigma_{b}^2)$.

The Bayesian model is completed with the elicitation of a prior distribution for the parameters and hyperparameters
 $\boldsymbol \theta =(\boldsymbol \beta_0, \boldsymbol \beta_W, \boldsymbol \beta_A, \sigma, \sigma_a, \sigma_b)^\prime$ of the model. We assume prior independence between them and select the  uniform distribution $\text{U}(0,10)$ for all standard deviation terms. The elicited marginal prior distribution for  $\beta_0^{(a)}$  and $\beta_0^{(c)}$ is $\text{U}(0,20)$ and  $\text{U}(0,10)$, respectively.  These uniform distributions are sufficiently large to cover generously the whole range of possible values of both parameters. A normal distribution $\text{N}(0,10^2)$ is selected for $\beta_0^{(b)}$ to allow the parameter to move freely between  a wide range of positive and negative values.

\subsection{Posterior distribution}\label{subsec31}

The  relevant quantities in the inferential process are the parametric vector $\boldsymbol \theta$ and the set of random effects associated to the individuals in the sample $\boldsymbol u=(\boldsymbol u_1, \ldots, \boldsymbol u_n)^{\prime}$, where $\boldsymbol u_i=(u_i^{(a)}, u_i^{(b)})$.  The  posterior distribution  $\pi(\boldsymbol \theta, \boldsymbol u \mid \mathcal D)$, where $\mathcal D$ represents the observed data, contains all the relevant information of the problem and it is usually the starting    point of all relevant inferences. It was  approximated by means of Markov Chain Monte Carlo (MCMC) simulation methods through   the JAGS software \citep{Plummer}. For each estimated model, we ran three parallel chains with 1,000,000 iterations and a burn-in of 1,000,000. Chains were also thinned by storing every 1,000th iteration to reduce autocorrelation in the sample. Convergence to the joint posterior distribution was guaranteed by visualising every autocorrelation function plot by means of \verb"mcmcplot" package for the R software and assuring an effective number of independent simulation draws greater than 100. For the sake of reproducibility we have generated a fictitious databank, which together with the R code for the analyses is available as supplementary material here \url{https://github.com/gcalvobayarri/intra_abdominal_volume_model.git}.

Table \ref{tab:posterior} summarizes $\pi(\boldsymbol \theta, \boldsymbol u \mid \mathcal D)$. The posterior mean of $\beta_0^{(a)}$ and $\beta_0^{(b)}$ provides an approximate overall assessment of the baseline values  of  $a_i$ and $b_i$  for male patients. In the case of the
   asymptotic value   $a_i$, it decreases   by about 0.344 in the female group (although  this estimation has a lot of uncertainty), and shows a slight positive trend with age. Differences between individuals are relevant as it can be seen from the estimation of the standard deviation of the random effect in $a_i$, 1.743. The parameter $b_i$ has an approximate basal value of 0.922 in the men group, which decreases by -0.24 units in the women group. Age also has a positive estimation and   the random effect associated to individuals are also important for $b_i$, especially because this term   appears on an exponential scale and negative sign in the quotient of the growth curve. Finally, the posterior mean for the $c_i$ term is about 2.184 in the men group and decreases in 0.245 units in the group of women. The posterior mean of the standard deviation associated to the measurement error  is  not very large but it does have a very high accuracy. The fact that   the $IP$, $ADP$, $MAP$ and $MDP$ of individual $i$  depends on $b_i$ and $c_i$ proportionally to $-b_i/c_i$ and that the estimated  coefficients associated to gender are negative  for both
   $b_i$ and $c_i$ implies that  $IP$'s, $ADP$'s, $MAP$'s and $MDP$'s for women will be slightly higher than the subsequent for men. The relationship of the $IP$, $ADP$, $MAP$  and $MDP$ with age is negative but barely important.

\begin{table}[h]
 \centering
\caption{Posterior summaries (mean, standard deviation and $95\%$ credible interval)  for the parameters and hyperparameters of the logistic   growth  model with covariates gender and  standardized   age.}\label{tab:posterior}
\vspace{0.25cm}
\begin{tabular}{crrc}
\noalign{\hrule height 1pt}
                & \multicolumn{3}{c}{Logistic growth model}                                                                            \\ \cline{2-4}
  Parameters    & \multicolumn{1}{c}{mean} & \multicolumn{1}{c}{sd} & \multicolumn{1}{c}{$IC_{0.95}$}    \\ \hline
$\beta_0^{(a)}$ & $5.597$                  & $0.392$                & $(4.861$, $6.376)$                         \\
$\beta_W^{(a)}$ & $-0.344$                 & $0.264$                & $(-0.875$, $0.153)$                         \\
$\beta_{Age}^{(a)}$ & $0.110$                  & $0.122$                & $(-0.123$,  $0.347)$                         \\\vspace*{0.12cm}
$\sigma_a$      & $1.743$                  & $0.095$                & $(1.571$, $1.938)$                         \\
$\beta_0^{(b)}$ & $0.922$                  & $0.166$                & $(0.601$, $1.238)$                         \\
$\beta_W^{(b)}$ & $-0.246$                 & $0.112$                & $(-0.464$, $-0.028)$                        \\
$\beta_{Age}^{(b)}$ & $0.120$                  & $0.054$              & $(0.017$, $0.224)$                         \\\vspace*{0.12cm}
$\sigma_b$      & $0.733$                  & $0.041$                & $(0.658$, $0.818)$                         \\
$\beta_0^{(c)}$ & $2.184$                  & $0.040$                & $(2.108$, $2.262)$                         \\\vspace*{0.12cm}
$\beta_W^{(c)}$ & $-0.245$                 & $0.029$                & $(-0.300$, $-0.188)$                         \\
$\sigma$        & $0.361$                  & $0.003$                & $(0.355$, $0.367)$                         \\
\noalign{\hrule height 1pt}
\end{tabular}
\end{table}

The posterior distribution   is  the starting point for the analysis of the different outcomes of interest in the study. In the following, we will present different results that may be useful to better understand the relationship between $IAV$ and $IAP$ at both the individual and population level and thus be able to answer the scientific questions raised by the study.
But first we would like to make a brief comment on the variable selection process discussed above for parameter $c_i$ of the growth model. In this context, we   considered different modelling approaches  for $c_i$ with regard to covariate  gender.  The Deviance Information Criterion \citep{Spiegelhalter} was used for  model comparison and according to this criterion  the best  model was the one with only the gender covariate and a common population term in parameter $c_i$ as stated before.

\begin{figure}[h]
\begin{center}
\includegraphics[width= 9.5cm]{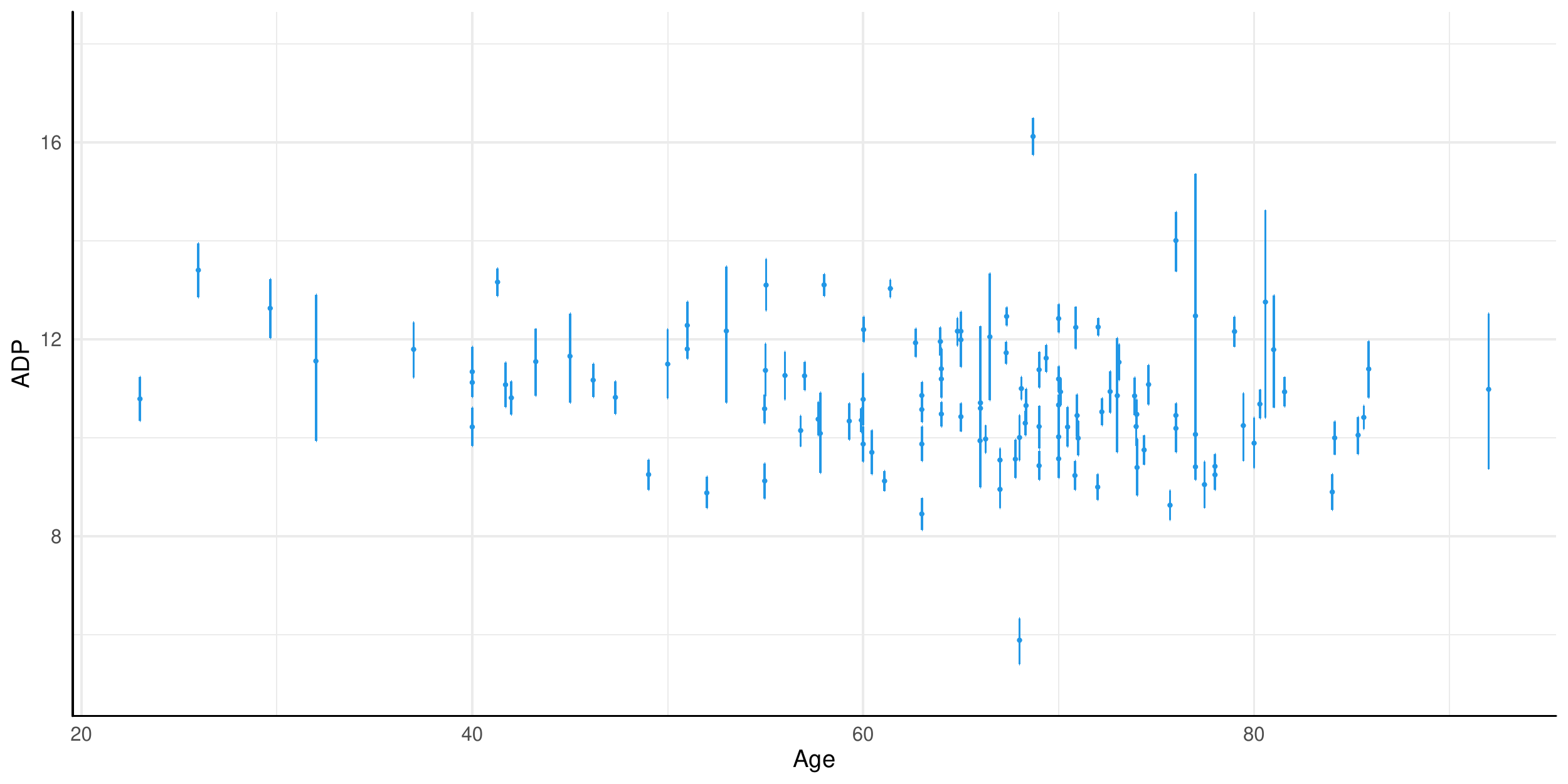} \includegraphics[width= 9.5cm]{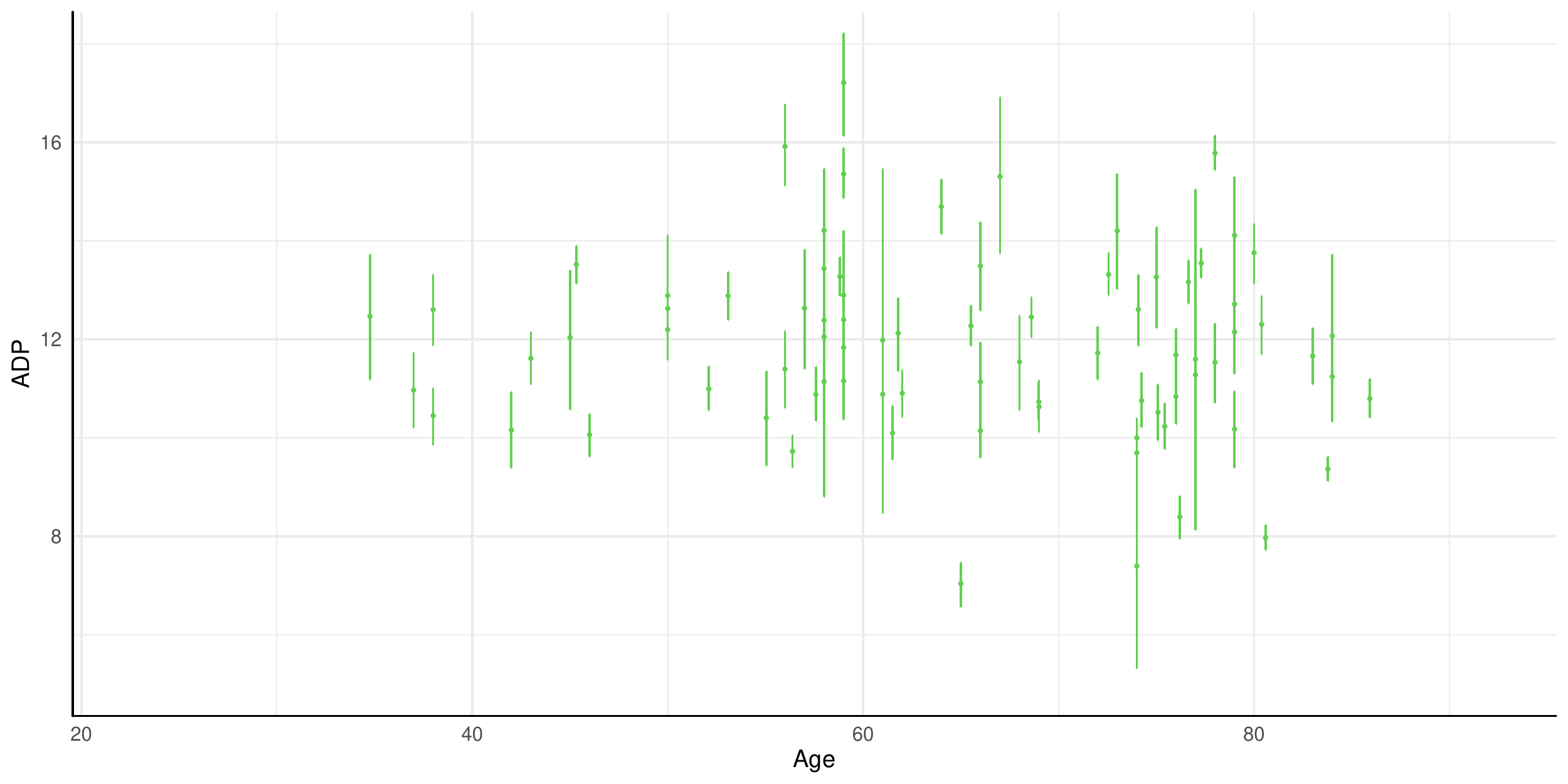}\\
\caption{Posterior mean and 95$\%$ credible interval  of the $ADP$ value of the  men  (top panel) and   the women   (bottom panel) in the sample.  Patients are ordered in the $x$-axis according to their age from youngest to oldest. } \label{fig:individual}
\end{center}
\end{figure}

\subsection{Posterior individual  outcomes}\label{subsec32}

The basic inferential process allows the Bayesian methodology to obtain information both individually and in terms of the target population.

In the following we focus on $ADP$. The  true $ADP$ value for  individual $i$, $ADP_i$, depends on $b_i$ and $c_i$, which in turn depend on  $(\boldsymbol \theta, \boldsymbol u_i)$. Consequently, we can compute the posterior distribution of the true $ADP_i$ of each individual $i$ in the sample from the subsequent posterior distribution
$\pi(\boldsymbol \theta, \boldsymbol u_i \mid \mathcal D)$. Figure \ref{fig:individual} shows the posterior mean and a 95$\%$ credible interval for that quantity of the individuals in the sample ranked by age.  The first thing that is striking in both graphs is the great difference in both the men and women groups in the range of credibility intervals, which is mainly explained by the differences in the number of repeated observations for each of them. This situation is more evident in the women's group due to the low  number of repeated measures per individual with regard the subsequent number in the men's group.

\begin{figure}[h]
\begin{center}
\includegraphics[width= 10cm]{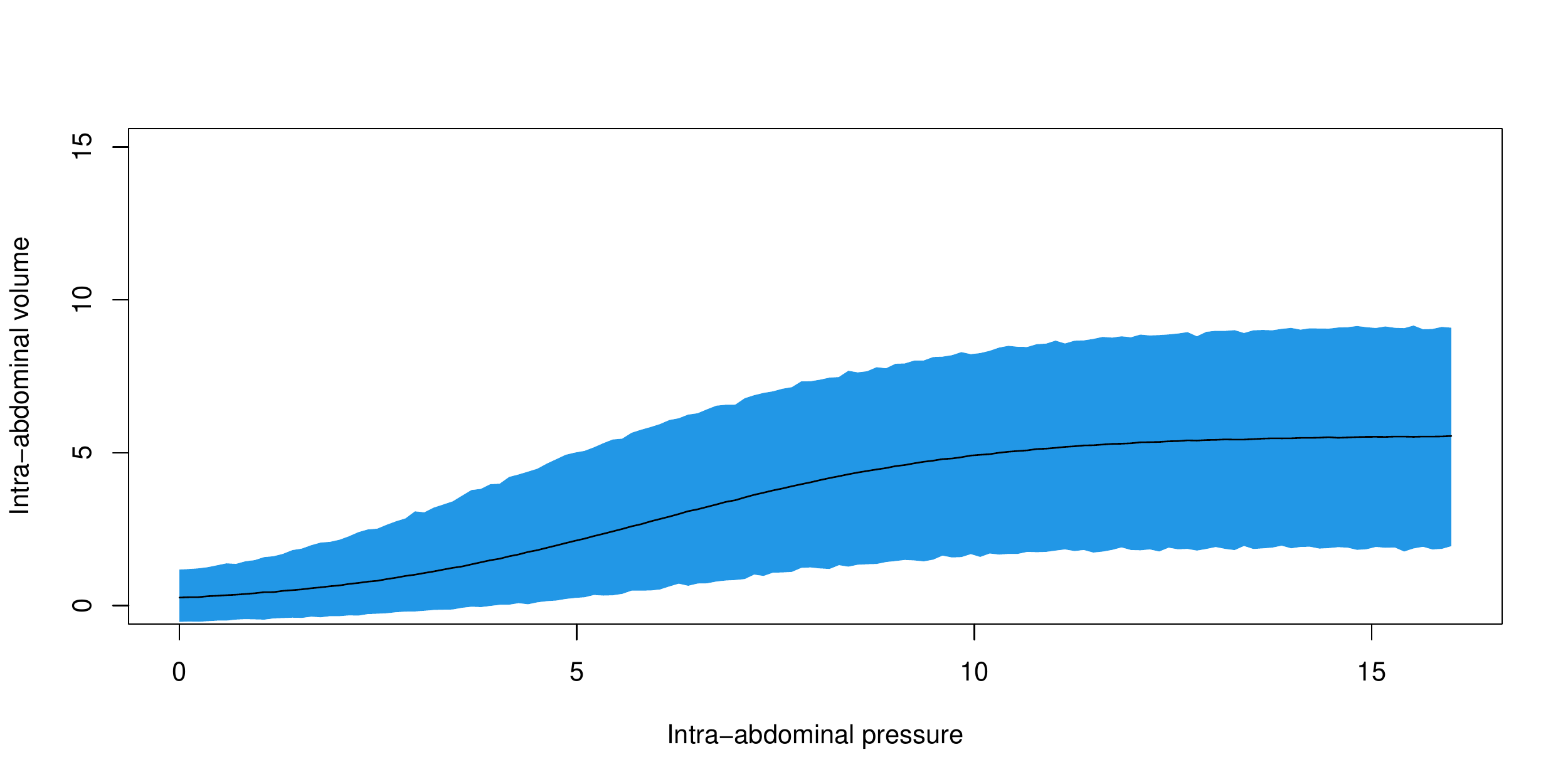}   \vspace*{-0.3cm}\\
\includegraphics[width= 10cm]{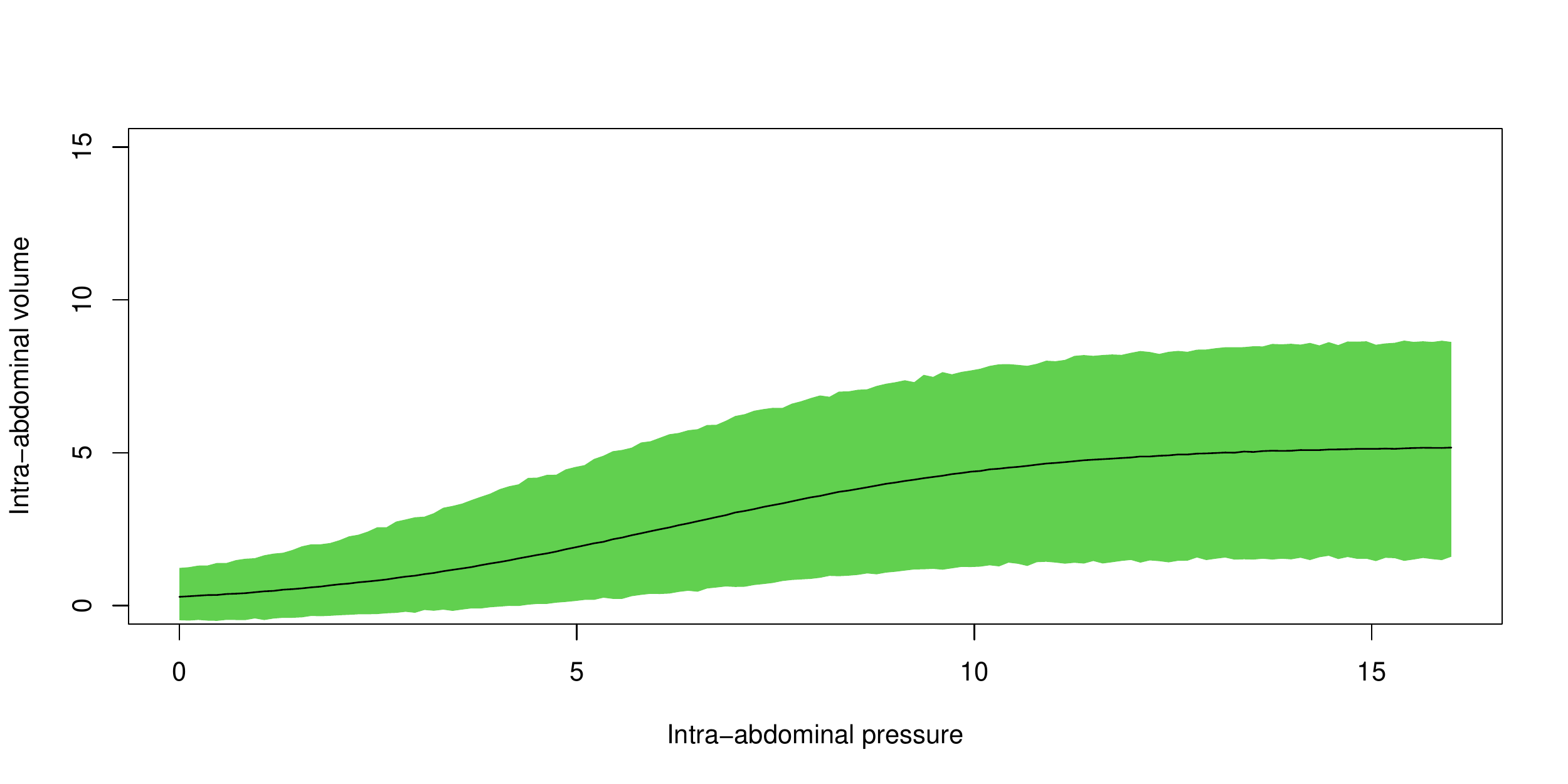}
\caption{Posterior predictive mean of the $IAV$ and 95$\%$ predictive interval with regard to $IAP$ values for a man (top panel) and  a woman  (bottom panel) aged 64.56 years (the sample mean).} \label{fig:predictive}
\end{center}
\end{figure}

The prediction of observations for   new individuals  in the target population is an important issue that  Bayesian statistics approaches in a natural way. The   posterior predictive distribution of the  random variable $Y_{n+1,j}$  that records the $IAV$  value for a new individual, $n+1$, of the population with regard to their $x_{n+1, j}$ values depends on the conditional model in (\ref{eqn:modela}) and the posterior distribution $\pi(\boldsymbol \theta, \boldsymbol u_{n+1} \mid  \mathcal D)$, where $\boldsymbol u_{n+1}$ are the random effects associated to that individual $n+1$, and is computed as follows

\begin{equation}
(Y_{n+1,j} \mid x_{n+1,j}, \mathcal D) \sim  \int \, (Y_{n+1,j} \mid  \boldsymbol \theta, \boldsymbol u_{n+1}) \, \pi(\boldsymbol \theta, \boldsymbol u_{n+1} \mid  \mathcal D)\, \mbox{d}(\boldsymbol \theta, \boldsymbol u_{n+1}),
\end{equation}

\noindent where the posterior $\pi(\boldsymbol \theta, u_{n+1} \mid  \mathcal D)$  factorizes in terms of the marginal posterior distribution $\pi(\boldsymbol \theta \mid \mathcal D)$ and the conditional  distributions for the   random effects $f(u^{a}_{n+1} \mid \sigma^2_{a})= N(0, \sigma^{2}_{a})$ and $f(u^{b}_{n+1} \mid \sigma^2_{b})= N(0, \sigma^{2}_{b})$. Figure \ref{fig:predictive} shows the posterior predictive mean and a 95$\%$ predictive interval for the $IAV$ value  of a new individual of the target population with respect to their $IAP$ and in relation to their gender. Both groups behave very similarly. The stabilisation of the values of $IAV$ in both groups can be clearly seen, as well as the variability associated with the predictive processes, which is always greater in comparison with the estimation processes themselves.

\subsection{Posterior population outcomes}\label{subsec33}

Random effects   connect    the different repeated measures of the same individual in the statistical model and allow for the computation of individual-specific outcomes. We would also like to be able to have not only that individual information, but also  outcomes   that can provide general information about the target population. This aim implies to work with the   marginal formulation of the model in (\ref{eqn:modela}) and (\ref{eqn:modelb}) which we would obtain by integrating out the random effects  as follows

 \begin{equation}
 (Y_{ij} \mid \boldsymbol \theta, x_{ij}) \sim \int\,  \mbox{N}(\mu_{ij}, \sigma^2)\, f(\boldsymbol u  \mid \boldsymbol \theta)\,\mbox{d}\boldsymbol u.
 \end{equation}

This marginal formulation   only depends on the parameter and hyperparameters of the model $\boldsymbol \theta$ and is the basis for the computation of  any feature of this marginal model. For simplicity, we only focus in the paper on the true asymptotic $IAV$ value and the true asymptotic deceleration point $ADP$ and its subsequent value for $IAV$.

\begin{figure}[h]
\begin{center}
\includegraphics[width= 6cm]{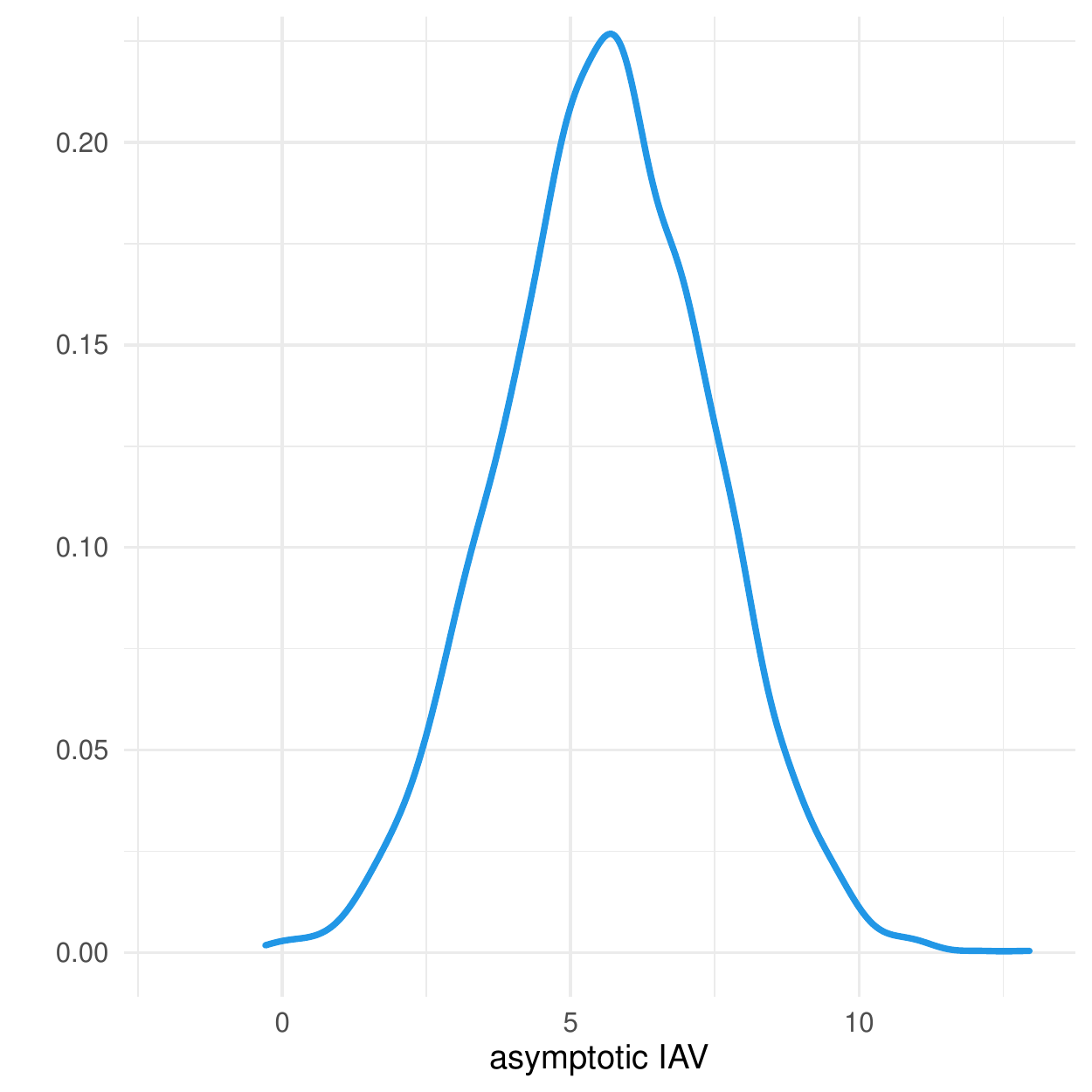}
\includegraphics[width= 6cm]{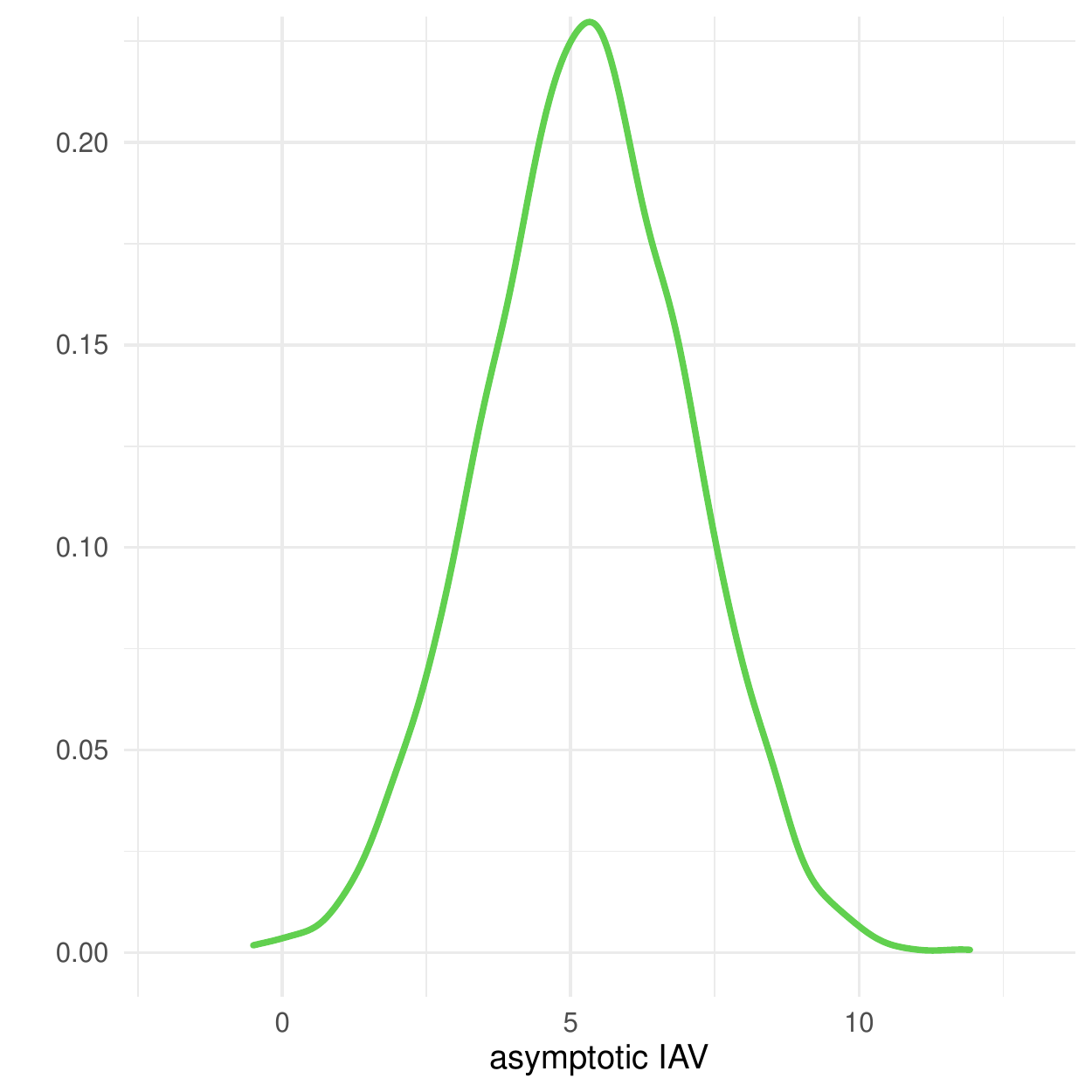}
\caption{Posterior distribution of the asymptotic  $IAV$   for  men  (on the left) and   women   (on the right).} \label{fig:apopulation}
\end{center}
\end{figure}
\begin{figure}[H]
\begin{center}
\includegraphics[width= 9cm]{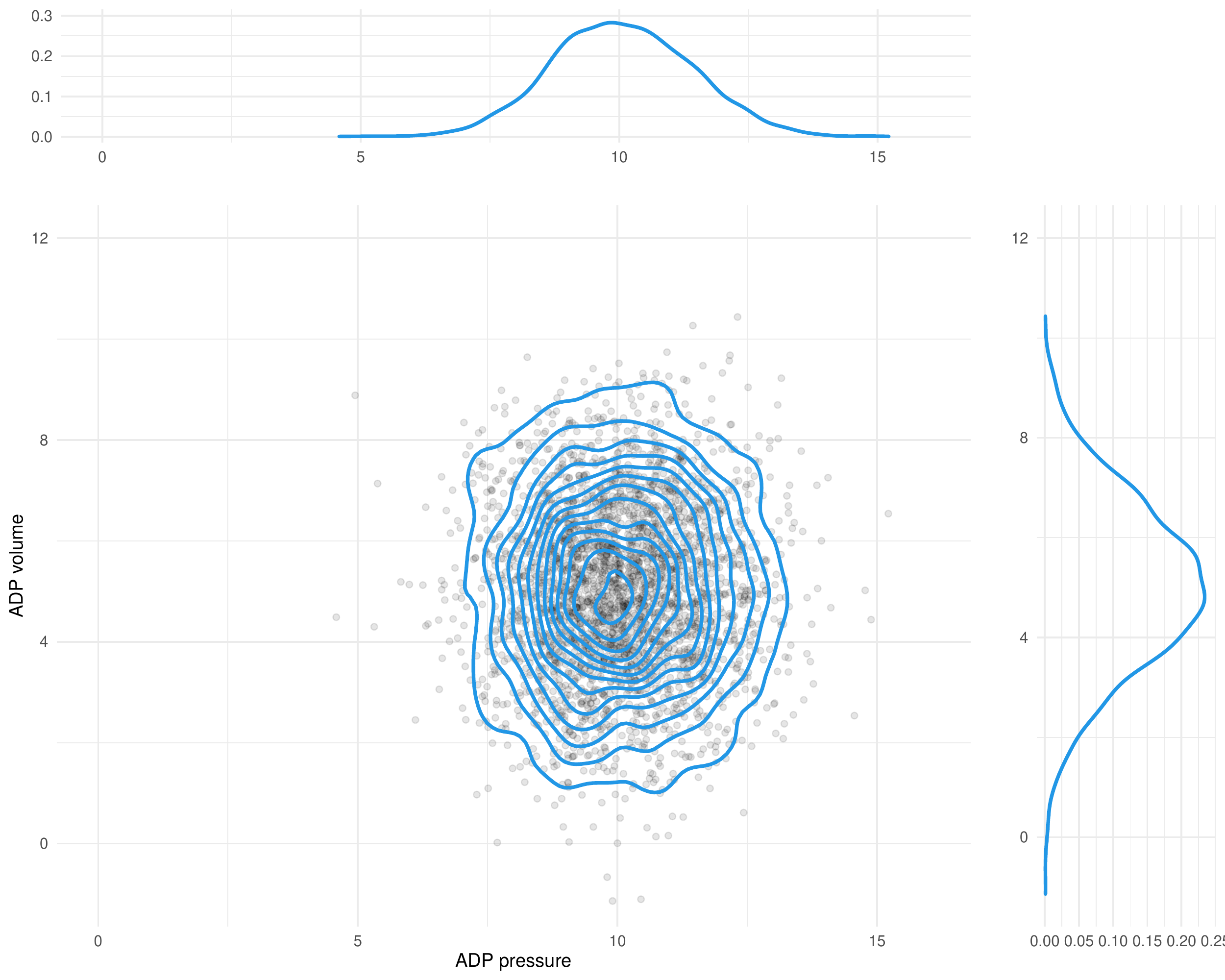}\\
\includegraphics[width= 9cm]{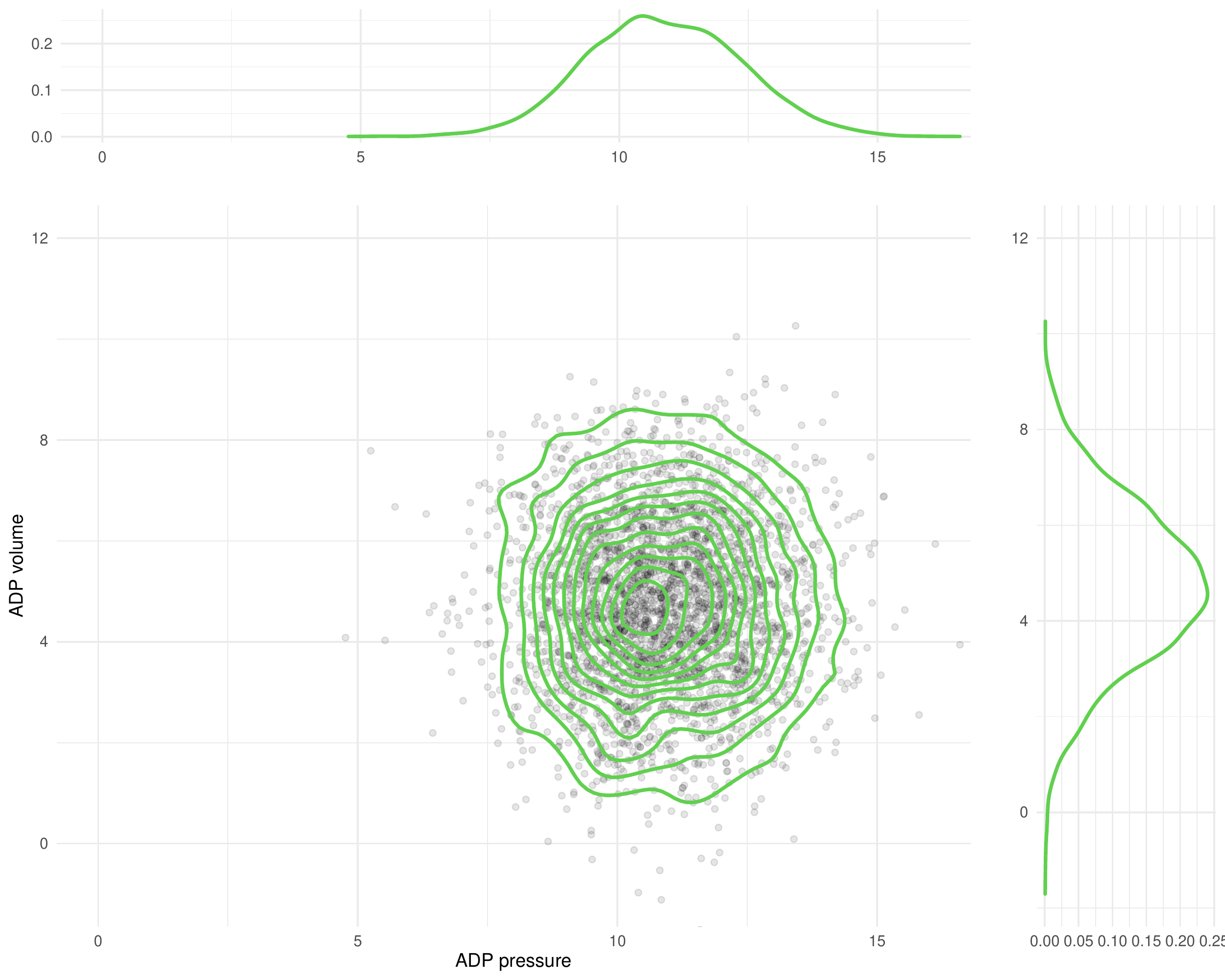}
\caption{Joint posterior distribution and contour lines  of the true $ADP$  and its subsequent  $IAV$ value and posterior marginal distribution for each of both quantities for  men  (top panel) and   women   (bottom panel) aged 64.56 years (the sample mean).} \label{fig:APDpopulation}
\end{center}
\end{figure}

Figure (\ref{fig:apopulation}) shows the posterior distribution  of the asymptotic $IAV$ for men and women aged 64.56 years (the mean of the sample). There is not much difference between the two distributions. An estimation of the asymptotic $IAV$ in the   group of men is 5.60 L. while in the  group of women  it is 5.25 L. Figure \ref{fig:APDpopulation} shows  the joint posterior distribution, in terms of contour lines, of the $ADP$ pressure point and the subsequent volume value for men and women aged 64.56 years (the sample mean) as well as the marginal distributions of both quantities. Posterior mean for the $ADP$'s pressure and volume is 10.06  mmHg.
 and 5.05 L. in men aged 64.56,  and 8.86  mmHg.
 and 4.12 L. in the group of women with the same age, respectively.

\section*{Conclusions}

Precision medicine tenets are that different interventions have distinct
effects in different people and that this variability can, at least in part, be
characterized and predicted \citep{Senn}. In this study we have tried to lay
the foundation for the mathematical modeling of the abdomen behavior during
pneumoperitoneum insufflation. We have also parameterized such model to achieve
predictive capability based on a few simple baseline characteristics. This is
the first step in a precision medicine approach to pneumoperitoneum
insufflation for laparoscopic surgery. This process can be potentially scaled
up and recursively performed throughout the duration of the surgical
intervention to ensure that even if conditions change, we could be able to
provide an optimal surgical field to the surgeon while exposing the patient the
lowest possible pressure.

With this procedure, we would like to achieve an optimal surgical workspace
while minimizing the pressure administered to the patient. In other words, each
subject would receive a titrated pressure according to his characteristics.
Also, the ability to predict where the marginal gain in volume diminishes by
deriving critical points on the parameterized curve have an especially
interesting clinical potential.

Bayesian inference can provide a suitable inferential framework in this context. First of all, Bayesian hierarchical models are useful to elicit and formulate the different sources of variation and uncertainty of the problem and incorporate suitable terms into the model to account for them. In this particular case, the model includes non-linear effects through a logistic growth function. As model fitting relies on MCMC methods, inference about particular elements of interest in the model becomes feasible. For example, the logistic growth model has a known parametric form from which some crucial critical points can be derived analytically but inference on these points is far from straightforward. However, the output produced by MCMC during model fitting can be exploited to compute the posterior marginals of these particular points as well as those of the other model parameters. This provides extra information that can be used  during the laparoscopic surgery. Inference about these critical points under other inferential frameworks would not be so straightforward.

The most important critical point in our study  is $ADP$, as this controls how much air is insufflated during surgery. From a clinical point of view, when operating on new patients, $ADP$'s predictive distribution can help physicians provide adequate insufflation during laparoscopic surgery. The study presented in this paper illustrates a preliminary analysis in which 198 patients have been enrolled. In the future, we aim to conduct a larger trial so that a wider range of patients is represented. Furthermore, other covariates will be recorded and included into the model to reduce the uncertainty about the estimates and predictions, and increase the accuracy of insufflation.

\section*{Acknowledgements}

This paper  was supported by research grant  PID2019-106341GB-I00  from Ministerio de
Ciencia e Innovaci\'on (Spain) and the Project MECESBAYES (SBPLY/17/180501/000491) from
the Consejer\'ia de Educaci\'on, Cultura y Deportes, Junta de Comunidades de Castilla-La Mancha
(Spain). Gabriel Calvo  is also supported    by grant    FPU18/03101 from the Ministerio de Ciencia e Innovaci\'on (MCI, Spain). Merck Sharp \& Dohme funded the IPPColLapse II study (Protocol Code No. 53607). This is an investigator-initiated study in which the sponsors and funders have no roles in study design,
analysis of data, or reporting.

\bibliographystyle{chicago}

\end{document}